\documentclass[aps,preprint,nofootinbib,preprintnumbers,eqsecnum,superscriptaddress]{revtex4}
\pdfoutput=1

\usepackage[
      colorlinks=true,
      linkcolor=blue,
      urlcolor=blue,
      filecolor=black,
      citecolor=red,
      pdfstartview=FitV,
      pdftitle={},
        pdfauthor={Alejandra Castro, Stephane Detournay, Nabil Iqbal, Eric Perlmutter},
        pdfsubject={},
        pdfkeywords={},
        pdfpagemode=None,
        bookmarksopen=true
      ]{hyperref}

\usepackage[normalem]{ulem}
\usepackage{enumerate}
\usepackage{amsfonts}
\usepackage{yfonts}

\usepackage{subfigure}
\usepackage{psfrag}

\usepackage{epsfig}
\usepackage[latin1]{inputenc}
\usepackage{float}
\usepackage{graphicx}
\usepackage{cancel}
\usepackage{mathrsfs}
\usepackage{amssymb}
\usepackage{amsfonts}
\usepackage{amsmath}
\usepackage{slashed}

\usepackage{graphicx}
\usepackage{latexsym,ifpdf}
\newcommand{\hc}{{\rm h.c.}}
\newcommand{\ft}[2]{{\textstyle\frac{#1}{#2}}}
\def\rmi{{\rm i}}
\def\rmd{{\rm d}}
\def\rme{{\rm e}}
\def\Re{\mathop{\rm Re}\nolimits}

\newcommand{\QED}{{\hspace*{\fill}\rule{2mm}{2mm}\linebreak}}

\newcommand{\SO}{\mathop{\rm SO}}
\newcommand{\U}{\mathop{\rm {}U}}

\newcommand{\OSp}{\mathop{\rm {}OSp}}

\newcommand{\su}{\mathop{\mathfrak{su}}}   
\newcommand{\soa}{\mathop{\mathfrak{so}}}   

\newtheorem{theorem}{Theorem}
\begin{document}

\title {Construction of the de Sitter supergravity}


\author{Eric Bergshoeff}
\affiliation{Van Swinderen Institute for Particle Physics and Gravity,\\
University of Groningen, Nijenborgh 4, 9747 AG Groningen, Netherlands}

\author{Dan Freedman}
\affiliation{SITP and Department of Physics, Stanford University, Stanford, California
94305 USA \\
Center for Theoretical Physics and Department of Mathematics, Massachusetts Institute of Technology, Cambridge, Massachusetts 02139, USA}

\author{Renata Kallosh}
\affiliation{SITP and Department of Physics, Stanford University, Stanford, California
94305 USA  }

\author{Antoine Van Proeyen}
\affiliation{KU Leuven, Institute for Theoretical Physics,\\
       Celestijnenlaan 200D, B-3001 Leuven, Belgium.}

\begin{abstract}
\vskip 0.3in
Recently, the complete action for an ${\cal N}=1$ pure supergravity action in 4 dimensions that allows a positive, negative or zero cosmological constant has been constructed. The action is the generalization of a Volkov-Akulov action for the Goldstino coupled to supergravity. The construction uses a nilpotent multiplet.

This paper is written in honour of Philippe Spindel. AVP enjoyed collaborations and many interactions with Philippe, who has always appreciated very precise derivations. We use this occasion to give a very detailed account of the calculations that lead to the published results. We review aspects of supersymmetry with de Sitter backgrounds, the treatment of auxiliary fields, and other ingredients in the construction.
\end{abstract}

\vfill


\maketitle

\tableofcontents

\section{Introduction}
The very first paper on supergravity and the cosmological constant was written by Townsend in January 1977 \cite{Townsend:1977qa}.
Using an iterative Noether procedure,  he showed that a cosmological constant could be added to the then one year old supergravity action consistent with local supersymmetry. Furthermore, he found that the sign of the cosmological constant $\Lambda$ was negative leading to an AdS background solution. More specifically, the Lagrangian was given by
\begin{equation}
\label{massterm}
{\cal L}_{\rm AdS} = \frac{e}{\kappa^2}\bigl( \tfrac{1}{2}R(e,\psi) -\tfrac{1}{2}{\bar \psi}_\mu\gamma^{\mu\nu\rho}D_\nu\psi_\rho
+\tfrac{1}{2}m{\bar\psi}_\mu\gamma^{\mu\nu}\psi_\nu +3m^2\big)\,,
\end{equation}
such that
\begin{equation}
\label{value}
\Lambda = -3\frac{m^2}{\kappa^2} < 0\,.
\end{equation}
Here $m$ is a mass parameter that occurs in a mass-like term for the gravitino and defines the value of the cosmological constant via \eqref{value}.
We note that the specific sign of $\Lambda$ was not stressed in \cite{Townsend:1977qa}. The main interest of this paper was to show that a cosmological constant consistent with local supersymmetry could be added in the first place. At that time it was not yet known that we are living in a universe with a small positive cosmological constant. It might seems strange that the Lagrangian \eqref{massterm} contains a mass term for the graviton and gravitino but this is misleading because in the presence of a cosmological constant the Minkowski spacetime is not a solution anymore. Instead, to determine the spectrum, we should expand around an AdS vacuum solution.
This has the effect that in the expansion of the Einstein-Hilbert term the ordinary derivatives of Minkowski space get replaced by covariant derivatives using the AdS background metric. The effect of these covariant derivatives is that they lead to an additional mass-like term for the graviton  that precisely cancels the explicit mass terms present in the Lagrangian \eqref{massterm}. Due to the presence of the local supersymmetry, precisely the same happens in the case of the gravitino such that we
end up with a massless supergravity multiplet for the fluctuations around the AdS background.

The paper of Townsend was written less than a year after the invention of supergravity. A few months later, in April 1977, there was another paper on supergravity and the cosmological constant but this time in the context of broken supersymmetry \cite{Deser:1977uq}. It was argued that a Lagrangian could exist where de Sitter solutions are possible with broken supersymmetry. However, the explicit action including fermionic terms in supergravity for actions with nilpotent chiral multiplets had not yet been obtained (though constructions in 2 and 3 dimensions were done in \cite{Dereli:1977qy,Dereli:1977yx,Dereli:1977yk}). In 1978, it was shown how supergravity can be formulated using auxiliary fields \cite{Ferrara:1978em,Stelle:1978ye,Fradkin:1978jq} and how this supergravity can be constructed using superconformal methods \cite{Kaku:1978ea}. 37 years later, we constructed the `de Sitter supergravity' theory, allowing de Sitter vacua, using these methods.
We reported about its construction in  \cite{Bergshoeff:2015tra}. The same result up to field redefinitions has been obtained in \cite{Hasegawa:2015bza} by using other gauge choices for the superconformal symmetries.

Nilpotent superfields are already known since a long time. They lead to non-linear actions of fermions of the type of Volkov-Akulov \cite{Volkov:1973ix}, which were related to constrained chiral superfields in \cite{Rocek:1978nb,Ivanov:1978mx,Lindstrom:1979kq,Casalbuoni:1988xh}. These were also considered in the context of supergravity in \cite{Samuel:1982uh} and related to Born-Infeld actions, as far as the bosonic terms are concerned,  in \cite{Cecotti:1986gb}. The multiplets give a nice description of (partially) broken supersymmetry as studied in \cite{Bagger:1996wp,Komargodski:2009rz}.

In this paper, we will explain the steps for the construction of de Sitter supergravity \cite{Bergshoeff:2015tra} in more detail.

In Sec. \ref{ss:algebrasdS} we will recapitulate what is the problem with super-de Sitter algebras, and what are the consequences thereof for de Sitter supergravity. Then we repeat the construction of the rigid supersymmetry action with a nilpotent chiral multiplet from \cite{Komargodski:2009rz}. We reformulate it in our conventions for easy comparison with future results. We encounter the problem of the integration of an auxiliary field that does not appear just quadratically in the action, and for which, therefore, the Gaussian integration cannot be applied. We prove a theorem that can be used in different actions in Sec. \ref{ss:solnF}. In Sec. \ref{ss:pureVAsugra} we give a detailed account of the steps from the superconformal action to the pure de Sitter supergravity, explaining the elimination of auxiliary fields, the gauge fixing, several simplifications and the transformation laws. We finish with some comments on the result and later work in Sec. \ref{ss:comments}. Three appendices explain more details.


\section{Superalgebras and de Sitter supergravity}
\label{ss:algebrasdS}

The de Sitter and anti-de Sitter algebras are characterized by the fact that translations do not commute, but instead have a non-zero commutator given by (with $\mu=0,1,\dots,D-1$)
\begin{equation}
  \left[ P_\mu ,\, P_\nu \right] = \mp \frac{1}{4L^2} M_{\mu \nu }\,,
 \label{dSAdS}
\end{equation}
where $L$ is a length scale. We use the notation \cite{Freedman:2012zz} where the Lorentz algebra is given by
\begin{equation}
  [M_{\mu \nu } , M_{\rho \sigma }]\,=\,4\eta_{[\mu [\rho } M_{\sigma ]\nu ]}=\eta_{\mu \rho } M_{\sigma \nu }-\eta_{\nu \rho } M_{\sigma \mu }
-\eta_{\mu \sigma  } M_{\rho \nu }+\eta_{\nu\sigma } M_{\rho  \mu }\,.
 \label{Lorentzalgebra}
\end{equation}
The upper and lower signs in (\ref{dSAdS}) correspond to a de Sitter and anti-de Sitter
algebra, respectively. The Lie algebras are $\SO(D,1)$ for de Sitter  and $\SO(D-1,2)$ for anti-de Sitter.

When one tries to embed this algebra in a superalgebra with one Majorana spinor generator \cite[Sec. 12.6.1]{Freedman:2012zz}, one finds that the Jacobi identities only allow such an extension for $D=4$ when the lower sign in (\ref{dSAdS}) is used, and thus this is an ${\cal N}=1$ super-AdS algebra. An ${\cal N}=1$ super-dS algebra does not exist.

Super-dS algebras do exist for even ${\cal N}$ in 4 dimensions (and for other $D\leq 6$). W. Nahm \cite{Nahm:1978tg} classified the superalgebras that have a de Sitter algebra as a factor for the bosonic subalgebra, and for $D=4$ they are $\OSp({\cal N}^*|2,2)$,  which have an $R$-symmetry algebra $\soa^*({\cal N})$. These exist only for even ${\cal N}$ and are $\soa^*(2)=\soa(1,1)$, $\soa^*(4)=\su(1,1)\oplus \su(2)$, $\soa^*(6)=\su(3,1)$, and $\soa^*(8)=\soa(6,2)$. For higher ${\cal N}$ the $\soa^*({\cal N})$ have a different form, but we do not expect that such Lagrangian theories can exist. Hence all these $R$-symmetry algebras are non-compact. This hints at ghosts, which has been made explicit in \cite{Pilch:1985aw,Lukierski:1985it}.

A supergravity theory has an algebra based on structure functions, and not on structure constants. These have been called `soft algebras'. Hence the previous analysis does not restrict supergravity constructions. A supergravity theory is called a `de Sitter supergravity' if there is a solution of the field equations where the metric takes the form of a de Sitter space (and thus has positive cosmological constant in that solution). Then the Killing vectors of that solution satisfy a de Sitter algebra. What the previous analysis does imply is that there is no supersymmetric extension (for ${\cal N}=1$, $D=4$) of this algebra of Killing vectors. In other words, the solution cannot preserve supersymmetry.

Thus, we conclude that there is a no-go theorem based on Jacobi identities, that a solution of the field equations of a supergravity theory cannot preserve a de Sitter algebra and supersymmetry at the same time. Supergravity theories can exist that have a de Sitter solution (background) and in that solution, supersymmetry is broken. We remark that it has been shown in \cite{Anous:2014lia} that  (classical) superconformal theories do exist in a de Sitter background, because the de Sitter spacetime is related to Minkowski space by a Weyl rescaling of the metric.

\section{The rigid supersymmetry model revisited}
\label{ss:rigidcase}

In order to illustrate our method we first rewrite the model of Sect. 3.1 of \cite{Komargodski:2009rz}. This is a model with rigid supersymmetry, with a chiral multiplet $X$ that satisfies the constraint $X^2=0$. Following \cite{Kuzenko:2011tj,Ferrara:2014kva}, we write the action using a Lagrange multiplier chiral multiplet $\Lambda $ as
\begin{equation}
  {\cal L}=[ X \bar X]_D +\left[f\,  X\right] _F+ \left[\Lambda  X^2\right] _F\,.
 \label{Lsymbtoy}
\end{equation}
We use the notations for the actions that $[C]_D$ and $[Z]_F$ refer respectively to
\begin{equation}
  [C]_D=\int \rmd^4x\, \ft12 D+\ldots \,,\qquad [Z]_F =\int \rmd^4x\, \left(F +\ldots + \bar F+\ldots \right)= \int \rmd^4x\, 2\Re F+\ldots \,,
 \label{DFnotation}
\end{equation}
where the $\ldots $ refer to extra terms that appear in supergravity, $D$ is the last component of the real multiplet that has $C$ as lowest component, and $F$ is the last component of the chiral multiplet that has $Z$ as lowest component. $X$ is the lowest component of the chiral multiplet that becomes nilpotent, and $\Lambda$ is the lowest component of the chiral Lagrange multiplier multiplet. The second term contains a real constant $f$, which in supergravity will be generated from the value of the compensating multiplet.

The Lagrangian (\ref{Lsymbtoy}) is supersymmetric since it is built from the well-known ingredients of multiplet calculus or superspace. Before the equation of motion for $\Lambda $ is solved, this is just linear supersymmetry.

We denote the components of the two chiral multiplets in (\ref{Lsymbtoy}) as
\begin{equation}
  (X,\chi,F)\,,\qquad (\Lambda, {\chi^\Lambda}, F^\Lambda)\,.
 \label{componentstoy}
\end{equation}
The full Lagrangian (\ref{Lsymbtoy}) can be written in components as
\begin{eqnarray}
 {\cal L} & = &{\cal L}_1(X,\chi,F)+{\cal L}_c(X,\chi,F,\Lambda, {\chi^\Lambda}, F^\Lambda)\,,\nonumber\\
  {\cal L}_1&=& -\partial  _\mu X {\partial }^\mu \bar X-\frac12\bar \chi
  {\slashed{\partial  }}\chi+F \bar F 
 +2f\, \Re F\,, \nonumber\\
{\cal L}_c&=&F^\Lambda\,X^2 +\Lambda\left( 2 X\,F-\bar \chi P_L\chi\right) -2\bar \chi^\Lambda  P_L\chi X+\hc\,.
 \label{fullactiontoy}
\end{eqnarray}
The field equations are as follows \cite{Kuzenko:2011tj}:
\begin{eqnarray}
 F^\Lambda & : & X^2=0\,, \label{feH}\\
 {\chi^\Lambda} & : & X\,P_L\chi=0\,,\label{fechi}\\
 \Lambda&:&2 X\,F-\bar \chi P_L\chi=0\,,\label{feLambda}\\
 F&:&\bar F + f +  2 \Lambda X =0\,,\label{feF}\\
P_L \chi&:&-P_L{\slashed{\partial  }}\chi -2X P_L{\chi^\Lambda}-2\Lambda P_L\chi=0\,,\label{feOmega}\\
X&:&  \Box \bar X + 2F^\Lambda\,X +2\Lambda \, F -2\bar \chi^\Lambda P_L\chi=0\,.\label{feX}
\end{eqnarray}
The first three equations contain the three components of the chiral multiplet $X^2$. With $f\neq 0$, the field $F$ will have a nonzero value, which allows us to solve  (\ref{feLambda}) by taking \begin{equation}
X(\chi,F)= \frac{\bar \chi P_L\chi}{2 \,F} \equiv \frac{\chi^2}{2 \,F}\,,
\label{XinOmegaF}
\end{equation}
where
\begin{equation}
  \chi^2\equiv \bar \chi P_L\chi \,,\qquad\bar \chi^2= \bar \chi P_R\chi \,.
 \label{defOmega2}
\end{equation}
Since $P_L\chi\overline{\chi}P_L \chi=0$ (see the argument after (6.26) in \cite{Freedman:2012zz}), this condition solves also the first two field equations:  (\ref{feH}) and  (\ref{fechi}).

We insert this solution already in the action, such that
the last line of (\ref{fullactiontoy}) vanishes and the action reduces to an effective action that is the first line with the replacement (\ref{XinOmegaF}). We write it as
\begin{equation}
  {\cal L}(\chi,F)={\cal L}_1(X(\chi,F),\chi,F)=\left(\frac{\chi^2}{2 \,F}\right)  \Box
\left( \frac{\bar \chi ^2}{ 2 \,\bar F}\right)-\frac12\bar \chi
  {\slashed{\partial  }}\chi+F \bar F  +2f\, \Re F\,.
  \label{effectiveL}
\end{equation}
In Appendix \ref{app:feLagrM} it is proven (in a more general context) that the field equations obtained with this effective Lagrangian are also those that follow from the field equations including the Lagrange multipliers, (\ref{feH}) -- (\ref{feX}).

We write the Lagrangian as
\begin{equation}
 {\cal L}(\chi,F)= (F+f) (\bar F+f) +\left(\frac{\chi^2}{2 \,F}\right)  \Box
\left( \frac{\bar \chi ^2}{ 2 \,\bar F}\right)-f^2 -\frac12\bar \chi
  {\slashed{\partial  }}\chi\,.
 \label{fullactiontoyO}
\end{equation}
Without the substitution of $X$ as in (\ref{XinOmegaF}), we could have eliminated the auxiliary field $F$ by a Gaussian integration, and the first term in (\ref{fullactiontoyO}) would have disappeared. However, the peculiar feature in this action is the non-linear appearance of the auxiliary field $F$. In the next section, we will show how the auxiliary field can be eliminated in actions of this form. The theorem that we prove there (see (\ref{startLthm}) and (\ref{endLthm}) with $A=\Box$ and $B=0$) leads to:
\begin{equation}
 {\cal L}(\chi,F(\chi ))= -\frac12\bar \chi
  {\slashed{\partial  }}\chi-f^2+ \frac1{4f^2}\bar \chi^2\Box \chi^2-\frac{1}{16f^6}\chi^2\bar \chi^2(\Box \chi^2)(\Box \bar \chi^2)\,.
 \label{LequivVA}
\end{equation}
This is the result obtained in \cite{Komargodski:2009rz}.
It is proven in \cite{Kuzenko:2010ef} that after nonlinear field redefinitions this action is equal to the Volkov-Akulov action \cite{Volkov:1973ix}.

The action (\ref{LequivVA}) is still invariant under the supersymmetry transformation
\begin{equation}
  \delta(\epsilon)  P_L\chi  = \frac1{\sqrt{2}} P_L(\slashed{\partial} X +
F)\epsilon\,,
 \label{susytrOmega}
\end{equation}
where $X$ is
\begin{equation}
  X=-\frac{\chi ^2}{2f}(1-{\cal A})\,,
 \label{valueXrigid}
\end{equation}
and here
\begin{equation}
  {\cal A}= \frac{1}{4f^4}\bar \chi ^2\Box\chi ^2\,.
 \label{calArigid}
\end{equation}
$F$ is substituted in (\ref{susytrOmega}) by (\ref{finalF2}). In this case
\begin{equation}
F(\chi )= -f\Big( 1+ \frac{\bar \chi^2 }{ 4 f^4}\Box \chi^2 -\frac{3}{ 16 f^8} \chi^2 \bar \chi^2 \Box \chi^2 \Box \bar \chi^2\Big)\,.
\label{Fvaluerigid}
\end{equation}

\section{Non-gaussian integration}
\label{ss:solnF}
We encounter Lagrangians of $X$ and $F$, which appear as follows:
\begin{equation}
e^{-1}{\cal L}(X,F) = (F+f)(\bar F +\bar f) - f\bar f + \bar X\, A\, X + X\bar B + B\bar X\,,
\label{action}
\end{equation}
where $f$ and $B$ are complex functions of the other fields. $A$ is an operator of second order of the form
\begin{equation}
  A= \Box + \rmi t^\mu \partial _\mu + \ft12\rmi e^{-1}\partial _\mu (e t^\mu) + r\,, \qquad
  \Box =\frac{1}{\sqrt{g}}\partial _\mu \sqrt{g}g^{\mu\nu}\partial _\nu\,,
 \label{formA}
\end{equation}
where $t^\mu$ and $r$ are real. The names $t^\mu $ and $r$ refer to the fact that they contain terms related to torsion and curvature, respectively.
This form is dictated by the fact that the action should be real modulo total derivatives. We have
\begin{equation}
  \int\rmd^4x\,e\, \bar X\, A\, X =\int\rmd^4x\,e\, X\,\bar A\, \bar X= \int\rmd^4x\,e \left[-\partial _\mu X\partial^\mu\bar X +\ft12 \rmi t^\mu\left(\bar X \partial _\mu X - X\partial _\mu\bar X \right)+rX\bar X\right]\,,
 \label{intXAX}
\end{equation}
where $\bar A$ has the $\rmi t^\mu $ terms with the opposite sign. For the rigid case, we have the simplifications (we use that $e=\sqrt{-g}$ is 1 for the rigid case):
\begin{equation}
 \mbox{Rigid case:}\ A=\Box=\partial _\mu\partial ^\mu\,,\qquad B=0\,,\qquad f=\bar f=\mbox{ constant}\,.
 \label{rigidvaluesABC}
\end{equation}

We will prove the following theorem (we neglect the factor $e^{-1}$ in (\ref{action}) since it plays no role here).
\begin{theorem}
The Lagrangian
\begin{equation}
  {\cal L}= (F+f)(\bar F +\bar f) + \bar X\, A\, X + X\bar B + B\bar X\,,\qquad  \mbox{where } X=\frac{\chi ^2}{2F}\,,
 \label{startLthm}
\end{equation}
reduces after using the equation of motion for $F$ and $\bar F$ to
\begin{equation}
  {\cal L}=  \bar Y\, A\, Y + Y\bar B + B\bar Y- \frac{|Y((AY+B))|^2}{f\bar f}\,,\qquad  \mbox{where } Y=-\frac{\chi ^2}{2f}\,.
 \label{endLthm}
\end{equation}
\end{theorem}

We consider the Lagrangian as depending on $F$ via two ways:
${\cal L}(F)={\cal L}(X(F),F)$,
where
$X(F) $ is the expression in (\ref{startLthm}) in terms of $F$ and the fermion square $\chi ^2=\bar \chi P_L\chi $. Then the fields $F$ and $\bar F$ can be eliminated using the algebraic eqs of motion:
\begin{equation}
\frac{\delta{\cal L}(F)}{\partial \bar F}=\frac{\delta{\cal L}(X,F)}{\partial \bar F}-\frac{\bar X}{\bar F}\frac{\delta{\cal L}(X,F)}{\partial \bar X}=0\  \ \longrightarrow \ \
F + f - \frac{\bar X}{\bar F} \left( A\, X + B\right)=0\,,
\label{Ffieldeqn}
\end{equation}
and its complex conjugate.
This implies first of all that
\begin{equation}
  F = -f+{\cal O}(\bar \chi^2)\,,\qquad \bar F = -\bar f+{\cal O}(\chi^2)\,,
 \label{Ffshort}
\end{equation}
where e.g. ${\cal O}(\bar \chi^2)$ means that the correction terms are proportional to an undifferentiated $\bar \chi^2$.
This then implies
\begin{equation}
  X= -\frac{\chi^2}{2f}+{\cal O}(\chi^2\bar \chi^2)= Y+{\cal O}(\chi^2\bar \chi^2)\,,
 \label{Xorderchi}
 \end{equation}
where $Y$ is of lowest order in $X$ as given in (\ref{endLthm}).
With (\ref{Ffieldeqn}) we obtain that the first term at the r.h.s.~of  (\ref{startLthm}) is given by
\begin{equation}
  (F+f)(\bar F +\bar f)= \frac{X\bar X}{F\bar F} | A\, X + B|^2\,.
 \label{Fplusf2}
\end{equation}
Due to the overall factor $X\bar X\propto \chi^2\bar \chi^2$, one can check that in the latter expression only the first terms of (\ref{Ffshort}) and  (\ref{Xorderchi}) contribute, and we can write
\begin{equation}
  (F+f)(\bar F +\bar f)= \frac{Y\bar Y}{f\bar f} | A\, Y + B|^2\,.
 \label{Fplusf2where}
\end{equation}
For future use we introduce the notation
\begin{equation}
  {\cal A}=\frac{\bar Y}{f\bar f}(AY+B)= \frac{\bar \chi^2}{ 2f \bar f^2} \left(A\, \frac{\chi^2}{2f} - B\right)\,.
 \label{defcalA}
\end{equation}
We write  (\ref{Fplusf2where}) as
\begin{equation}
  (F+f)(\bar F +\bar f)= f\bar f{\cal A}\overline{{\cal A}}\,.
 \label{FplusfcalA}
\end{equation}
The complete expression for $F$ is
\begin{equation}
 F = -  f \left[ 1-   \frac{\bar X}{ f \bar F} \left( A\,  X +  B\right)\right]\,.
 \label{FexpressionX}
\end{equation}
Since $\bar X$ is nilpotent, we have also
\begin{equation}
F^{-1} = -\frac{1}{f} \left[ 1+   \frac{\bar X}{ f \bar F} \left(A\, X + B\right)\right]\,.
\label{barF-1}
\end{equation}
This allows us to write the following expression for $X$:
\begin{equation}
X = \frac{\chi^2}{ 2}F^{-1}  =  - \frac{\chi^2}{ 2 f}  \left[ 1+   \frac{\bar X}{ f \bar F} \left(A\, X + B\right)\right]=Y\left(1-{\cal A}\right)\,.
\label{XinA}
\end{equation}
Indeed, the higher order terms of the second factor vanish in view of the $\chi^2\bar\chi^2$ overall factor. Observe that the two derivatives in $A$ must both act on $\chi^2$ in order not to be killed by the overall factor $\chi^2$.

Using (\ref{XinA}) the $X$-dependent terms in (\ref{startLthm}) are
\begin{eqnarray}
 \bar X\, A\, X &=& \bar Y A Y  -\bar Y\overline{{\cal A}}\, A\, Y -  \bar Y A( Y{\cal A}) + \bar Y \overline{{\cal A}} A( Y{\cal A})\,,\nonumber\\
X\bar B + B\bar X   & = & Y\bar B + B\bar Y -  Y{\cal A}\bar B + B\bar Y\overline{{\cal A}} \,.
 \label{Xtermsexpanded}
\end{eqnarray}
Since $Y{\cal A}$ is of order $\chi ^2\bar \chi ^2$, the last term of the first line vanishes. The third term, however, needs some more care. First of all, the $A$ operator should act on the two $\bar \chi $ factors in ${\cal A}$ in order that they are not killed by the factor $\bar Y$. Thus
\begin{equation}
  \bar Y A(Y{\cal A})= Y\bar YA{\cal A}\,,
 \label{step1calA}
\end{equation}
and $\bar YA{\cal A}$ is effectively of the form $\bar Y\partial \partial {\cal A}$. Since $\bar Y \partial {\cal A}=0$, we can 'integrate by parts' the two derivatives and obtain
\begin{equation}
  \bar YA{\cal A}= (A\bar Y){\cal A}\,.
 \label{step2calA}
\end{equation}
Therefore
\begin{eqnarray}
  \bar X\, A\, X + X\bar B + B\bar X &=& \bar Y\, A\, Y\phantom{ Y -  Y (A\bar Y){\cal A} + } + Y\bar B + B\bar Y \nonumber\\
  &&  -\bar Y\overline{{\cal A}}\, A\, Y -  Y (A\bar Y){\cal A} + Y{\cal A}\bar B + B\bar Y\overline{{\cal A}}\,.
 \label{correctionterm2}
\end{eqnarray}
Using the definition (\ref{defcalA}), the second line can be written as $-2f\bar f{\cal A}\overline{{\cal A}}$, and hence combines with (\ref{FplusfcalA}) to
$-f\bar f{\cal A}\overline{{\cal A}}$, which is the last term in (\ref{endLthm}). This proves the theorem.

\QED

This mechanism of flipping the sign of the $(F+f)(\bar F+\bar f)$ term, can be understood by defining ${\cal F}$ replacing the auxiliary field $F$:
\begin{equation}
  {\cal F}= F+f\,.
 \label{defcalF}
\end{equation}
After one has realized that ${\cal F}$ is of order $\bar \chi ^2$, one sees that
\begin{equation}
  X= \frac{\chi ^2}{-2f+2{\cal F}}=Y - Y\frac{{\cal F}}{f}\,,
 \label{XincalF}
\end{equation}
where ${\cal F}$ is the unknown part of the auxiliary field. The Lagrangian (\ref{startLthm}) can be written as (using the same steps as in the previous proof)
\begin{eqnarray}
  {\cal L}&=& \bar Y\, A\, Y + Y\bar B + B\bar Y +\nonumber\\
  &&+{\cal F}\overline{{\cal F}} - \frac{\overline{{\cal F}}}{\bar f}\bar YAY -   \frac{{\cal F}}{f}YA\bar Y - \frac{{\cal F}}{f}\bar B - \frac{\overline{{\cal F}}}{\bar f} B\bar Y\,.
 \label{startLthmYcalF}
\end{eqnarray}
The value of ${\cal F}$ up to the order that we need, is the same as if we consider it here as an independent field, with field equation
\begin{equation}
  {\cal F}= f{\cal A}= \frac{\bar Y}{\bar f}(AY+B)\,,
 \label{fecalF}
\end{equation}
and it is eliminated as in a Gaussian integration. Note, however, that we already use properties of the solution (proportionality of ${\cal F}$ to $\bar \chi ^2$) to obtain (\ref{startLthmYcalF}), and therefore one cannot really consider ${\cal F}$ as the auxiliary field. In fact  (\ref{fecalF}) does not give the full solution of $F$ that follows from (\ref{FexpressionX}). This solution, which one needs for supersymmetry transformation rules, was obtained in \cite{Bergshoeff:2015tra} as
\begin{equation}
  F= -  f\left[1+{\cal A} \left(1-3\bar{{\cal A}}-\frac{\chi^2}{2f^2\bar f}\bar B\right)\right]\,.
 \label{finalF2}
\end{equation}

\section{Pure VA supergravity}
\label{ss:pureVAsugra}

\subsection{Superconformal action}
In order to produce supergravity, we use the superconformal calculus, which implies that we need a compensating chiral multiplet. Its components are a scalar  $X^0$, a fermion $\chi ^0$ and an auxiliary field $F^0$. These are added to the nilpotent chiral multiplet that was used already in Sec. \ref{ss:rigidcase}, and that is now denoted as  $\{ X^1  ,\chi^1  ,F^1  \}$. Furthermore there is the Lagrange multiplier multiplet whose first component is $\Lambda$, see (\ref{componentstoy}).
All these multiplets have conformal weight~1.
We construct the action that corresponds to the two chiral multiplets $X^I= (X^{0}, X^1 )$ and the compensating multiplet $\Lambda $ in the form \cite{Ferrara:2014kva}
\begin{equation}
{\cal L} =  [N (X,\bar X)]_D + [\mathcal{W}(X)]_F + \left[\Lambda  (X^1 )^2\right] _F\,.
\label{symbL}
\end{equation}
In order to use the superconformal density formulas as in (\ref{DFnotation}), the expression for the $D$-term should have Weyl weight 2, and the $F$-terms should have Weyl weight~3.
We take the $U(1,1)$ invariant model, with $I=\{0,1 \}$, and $\eta_{IJ}$ diagonal with $\eta_{00}=-1$ and $\eta_{1 1 }=1$:
\begin{equation}
  N=\eta _{IJ}X^I\bar X^J=-X^0\bar X^0 +X^1   \bar X^1 \,,\qquad \mathcal{W}= a\left(\frac{X^0}{\sqrt{3}}\right)^3 + b \left(\frac{X^0}{\sqrt{3}}\right)^2 X^1 \,,
 \label{NWchoice}
\end{equation}
where $a$ and $b $ are dimensionless (not necessarily real) constants, which will provide respectively an anti-de Sitter supergravity and an uplifting of the potential to de Sitter.

In order to write the $D$ terms in (\ref{symbL}), we use the relation that for a chiral multiplet $(X,P_L\chi,F)$ of Weyl weight~1, the $D$-action can be written in the form of an $F$-action:
\begin{equation}
  [X\bar X]_D = \ft12 [X\bar F]_F\,,
 \label{DbecomesF}
\end{equation}
where $\bar F$ is the lowest component of a chiral multiplet of Weyl weight~2 since it transforms only under $P_L\epsilon $. The components of this multiplet are given in \cite[(16.36)]{Freedman:2012zz}:
\begin{equation}
  (\bar F, \slashed{\cal D}P_R\chi, \Box^C\bar X)\,.
 \label{barFmultiplet}
\end{equation}
The explicit expression of the superconformal covariant derivative is given in \cite[(16.34)]{Freedman:2012zz} and of the superconformal d'Alembertian on a scalar field of Weyl weight~1 in \cite[(16.37)]{Freedman:2012zz}.
These steps can be performed separately for the $X^0$ multiplet and for the $X^1 $ multiplet. Therefore, we write the Lagrangian as
\begin{equation}
  {\cal L} =  [\ft12 \eta_{IJ}X^I\bar F^J]_F + [\mathcal{W}(X^I)]_F + \left[\Lambda  (X^1 )^2\right] _F\,.
 \label{LinFformat}
\end{equation}
The component expression of the superconformal $F$-type action (\ref{LinFformat}) is given in \cite[(16.35)]{Freedman:2012zz}. The first term of (\ref{LinFformat}) is identical to \cite[(16.39)]{Freedman:2012zz}, where pure gravity was explained, and the ${\cal W}$ term was written in \cite[(17.19)]{Freedman:2012zz}. The full Lagrangian is thus
\begin{eqnarray}
{\cal L} & = &{\cal L}_1 + {\cal L}_c\nonumber\\
e^{-1}{\cal L}_1&=&\ft12\eta_{IJ}\left(F^I\bar F^J + X^I\Box^C \bar X^J -\bar \chi^I P_L\slashed{\cal D}\chi^J\right)\nonumber\\
&& +{\cal W}_I F^I -\ft12{\cal  W}_{IJ}\bar \chi^{I}P_L\chi ^{J}\nonumber\\
  &&+\frac{1}{\sqrt{2}}\bar \psi_\mu \gamma
  ^\mu \left[\ft12\eta_{IJ}\left(P_L\chi^I \bar F^J + X^I\slashed{\cal D}P_R\chi^J\right)+{\cal  W}_I P_L \chi^I\right]\nonumber\\
&&+\ft12 \bar \psi _{\mu }P_R \gamma ^{\mu \nu }\psi _{\nu }\left(\ft12\eta_{IJ}X^I\bar F^J+{\cal W}  \right)\nonumber\\
  &  & +\hc\,,\nonumber\\
e^{-1}{\cal L}_c &=&F^\Lambda\,(X^1 )^2 +  \Lambda \left(2X^1 F^1 -\bar \chi^1  P_L\chi^1 \right)\, -2\bar \chi^\Lambda  P_L\chi^1 X^1 \nonumber\\
&&+\frac{1}{\sqrt{2}}\bar \psi_\mu \gamma^\mu \left(2\Lambda X^1 P_L\chi^1 + (X^1 )^2P_L{\chi^\Lambda} \right)\nonumber\\
&&+\ft12 \bar \psi _{\mu }P_R \gamma ^{\mu \nu }\psi _{\nu }\Lambda (X^1 )^2  \nonumber\\
  &  & +\hc\,.
 \label{fullactionFform}
\end{eqnarray}
In these expressions appear the following derivatives of the superpotential in (\ref{NWchoice}):
\begin{eqnarray}
 \mathcal{W}_0 & = & 3 a \frac{(X^0)^2}{(\sqrt{3})^3} +\frac23b X^0 X^1 \,,\qquad \mathcal{W}_{1 }= \frac13b  (X^0)^2\,,\nonumber\\
 \mathcal{W}_{00}&=& 6 a \frac{X^0}{(\sqrt{3})^3} +\frac23b  X^1 \,,\qquad \mathcal{W}_{01 }=\frac23 b  X^0
  \,.
 \label{Wderiv}
\end{eqnarray}

The field equation of $\Lambda$ is
\begin{equation}
   2X^1 F^1 -\bar \chi^1  P_L\chi^1  +\sqrt{2}\bar \psi_\mu \gamma^\mu X^1 P_L\chi^1 +\ft12 \bar \psi _{\mu }P_R \gamma ^{\mu \nu }\psi _{\nu }(X^1 )^2 =0\,.
 \label{fieldeqnLambda}
\end{equation}
This is solved as in the rigid case by
\begin{equation}
  X^1 =\frac{\overline{\chi^1 }P_L \chi^1  }{2F^1 }\,,
 \label{Xnsolved}
\end{equation}
since this kills all components of the chiral multiplet $(X^1 )^2$.

\subsection{Simplifications of the action}

We want to eliminate the auxiliary fields, and use  (\ref{Xnsolved}). For the elimination of $F^1$, we will use the theorem of Sec. \ref{ss:solnF}. In order to write the action in the form (\ref{action}), it is convenient to write the $\bar X^1$ field equation. This will define what is $AX^1 + B$ in the terminology of (\ref{action}).
We will make use of the fact that this should also be a covariant equation modulo other field equations. This saves a lot of work, since we know that all the gauge connections recombine in covariant derivatives:
\begin{eqnarray}
  e^{-1}\frac{\delta S_1}{\delta \bar X^1 }&=&\Box^C X^1 +\overline{{\cal W}}_{01 }\bar F^0-\ft12\overline{{\cal W}}_{001 }\bar \chi^{0}P_R\chi ^{0}\nonumber\\
  &&+\frac{1}{\sqrt{2}}\bar \psi_\mu\gamma^\mu\left[\slashed{\cal D}P_L\chi^1 +\overline{{\cal W}}_{01 }P_R\chi^0+\frac{1}{\sqrt{2}}(F^1 +\overline{{\cal W}}_{1 })P_R\gamma^\nu\psi_\nu\right]\nonumber\\
  && -\ft12\bar \psi_\mu P_L\psi^\mu\left[ F^1 +\overline{{\cal W}}_{1 }\right]\,,
 \label{XL1fe}
\end{eqnarray}
where $S_1=\int \rmd^4x\,{\cal L}_1$. Note that the expression in square brackets in the second line is the field equation of $P_R\chi^1 $, while the one in the third line is the field equation of $\bar F^1 $.
Writing out some covariant derivatives leads to simplifications. One of these is that terms with $F^1$ all cancel. The simplification amounts to :
\begin{eqnarray}
  e^{-1}\frac{\delta S_1}{\delta \bar X^1 }&=&\Box'^C X^1 +\overline{{\cal W}_{01 }}\bar F^0-\ft12\overline{{\cal W}_{001 }}\bar \chi^{0}P_R\chi ^{0}\nonumber\\
  &&+\frac{1}{\sqrt{2}}\bar \psi_\mu\gamma^\mu \overline{{\cal W}}_{01 }P_R\chi^0
 + \frac{1}{\sqrt{2}}\bar \psi_\mu\gamma^{\mu\nu}P_L{\cal D}_\nu'\chi^1
 + \frac{1}{2}\bar \psi_\mu\gamma^{\mu\nu}P_L\psi_\nu\overline{{\cal W}}_{1 }\,,
  \label{XL1fesimpler}
\end{eqnarray}
where (valid for $\{X,\,\chi \}$ being $\{X^I,\, \chi ^I\}$, and used here for $\{X^1,\, \chi ^1\}$)
\begin{eqnarray}
 \Box'^C X  & = & e^{a\mu}\left(\partial_\mu {\cal D}_a  X-2 b_\mu {\cal D}_a X +\omega_{\mu\,ab}{\cal
D}^b  X +2f_{\mu a}X+\rmi A_\mu {\cal D}_a  X  +\frac{1}{\sqrt{2}} \bar \phi _\mu \gamma
_aP_L\chi\right)\,,\nonumber\\
   {\cal D}_a  X &=& e _a^\mu \left(\partial _\mu   X-\,b_\mu X -\rmi \, A_\mu
   X -\frac{1}{\sqrt{2}}\bar \psi _\mu P_L\chi\right)\,, \nonumber\\
 P_L {\cal D}'_\mu\chi &=& P_L\left[ \left( \partial _\mu+\frac14\omega _\mu {}^{bc}\gamma _{bc}
- \ft32 b_\mu   +\ft12\rmi A_\mu\right)\chi-\frac1{\sqrt{2}} \left(\slashed{\cal D} X \right)\psi _\mu  -\sqrt{2} X\phi _\mu \right]
\,.
\label{covderchiralmult}
\end{eqnarray}
There are further simplifications. E.g. all $b_\mu $ terms drop out if extracting them also from the spin connection $\omega _\mu {}^{ab}=\omega _\mu {}^{ab}(e,b,\psi )$ and correspondingly from $f^\mu _\mu $. This is a consequence of the special conformal invariance, since $b_\mu $ is the only field in this action that transforms under these transformations.

In Appendix \ref{app:detailsaction} we obtained an expression for (\ref{DbecomesF}) in a form convenient for  (\ref{action}), which we now use to write
\begin{eqnarray}
 {\cal L}_1
 & = &\eta _{IJ} \bar X^I  \left[\partial _\mu \sqrt{g}g^{\mu\nu}\partial _\nu + \rmi\,e\, t^\mu_c \partial _\mu + \ft12\rmi\,\partial _\mu (e\,t^\mu_c) + e\,r_0^c\right] X^J
  \nonumber\\
   &   & +e\, \eta _{IJ}\bar X^I\, B_c^J + e\,\eta _{IJ}X^I\bar B^J_c +e \, C_0^c+ e\,{\cal L}_{1,F}+e\,{\cal L}_{W,\rm ferm}\,,
 \label{12XbarXD}
\end{eqnarray}
(up to total derivatives). The indices $I=0,1$ and the subscript $c$ are a reminder that we are still in the superconformal setting with local conformal
symmetry (and other symmetries) unbroken:
\begin{eqnarray}
t^\mu_c &=&-2 A^\mu +\ft14 {\rmi} {\bar \psi}_\nu \gamma_\star \gamma^{\nu\rho\mu}\psi_\rho\,,\nonumber\\
r_0 ^c&=&-\ft16 R(\omega (e )) +\ft16\bar \psi _\mu \gamma ^{\mu \nu \rho } D^{(0)}_\nu \psi _\rho
  -A^\mu A_\mu -\ft16{\cal L}_{\rm SG,torsion} \,,\nonumber\\
B^I_c &=& \frac{1}{\sqrt{2}}\left[-e^{-1}\partial _\mu \left(e\bar \psi _\nu \gamma ^\mu \gamma ^\nu P_L  \chi^I\right)-\ft23\bar \chi^I P_L \gamma ^{\mu \nu }D _\mu \psi _\nu +\rmi A^\mu \bar \psi _\mu P_L\chi^I\right]\,, \nonumber\\
  C_0^c&=&\eta _{IJ} \left(  -\ft12\bar \chi^I  \slashed{D}^{(0)}\chi^J+\ft14\rmi \bar \chi^I \gamma _*\gamma ^\mu \chi^J A_\mu \right.\nonumber\\
 &&\phantom{\eta _{IJ}\ }  \left.-\ft{1}{32}\rmi\,e^{-1}\varepsilon ^{\mu \nu \rho \sigma }\bar \psi _\mu \gamma _\nu \psi _\rho \bar \chi^I \gamma _*\gamma _\sigma \chi^J-\ft12\bar \psi _\mu P_R\chi^I \bar \psi ^\mu P_L\chi^J \right) \,,\nonumber\\
 {\cal L}_{1,F} & = & \eta_{IJ}F^I\bar F^J+{\cal W}_IF^I +\overline{{\cal W}}_{\bar I}\bar  F^I\,,\nonumber\\
 {\cal L}_{W,\rm ferm}&=&-\ft12{\cal  W}_{IJ}\bar \chi^{I}P_L\chi ^{J}+\frac{1}{\sqrt{2}}\bar \psi_\mu \gamma
  ^\mu {\cal  W}_I P_L \chi^I+\ft12 \bar \psi _{\mu }P_R \gamma ^{\mu \nu }\psi _{\nu }\,{\cal W} +\hc \,.
 \label{12tr0C0}
\end{eqnarray}
Furthermore:
\begin{eqnarray}
  {\cal L}_{\rm SG,torsion}&=&-\ft{1}{16}\left[
(\bar{\psi}^\rho\gamma^\mu\psi^\nu) ( \bar{\psi}_\rho\gamma_\mu\psi_\nu
+2 \bar{\psi}_\rho\gamma_\nu\psi_\mu) - 4 (\bar{\psi}_\mu
\gamma\cdot\psi)(\bar{\psi}^\mu \gamma\cdot\psi)\right]\nonumber\\
&&-e^{-1}\partial _\mu \left(e\bar \psi \cdot \gamma \psi ^\mu \right)\,,\nonumber\\
D_\mu \psi _\nu &=&\left(\partial _\mu +\ft14 \omega _\mu {}^{ab}(e,\psi )\gamma _{ab}\right) \psi _\nu\,,\nonumber\\
 D^{(0)}_\mu   & = &\partial _\mu +\ft14 \omega _\mu {}^{ab}(e )\gamma _{ab} \,.
 \label{covder0}
\end{eqnarray}

Note that at this point one can recover the formulae of the rigid case described in Sec.~\ref{ss:rigidcase} by taking
$e_\mu{}^a = \delta_\mu{}^a, \psi_\mu=A_\mu=F^0=0$ together with $X^0=1, \chi^0=0$.

\subsection{Elimination of  auxiliary field \texorpdfstring{$F^0$}{F}}
\label{ss:elimauxf}
As mentioned before, the elimination of $F^1$ needs the theorem of non-Gaussian integration. However, we first already eliminate the auxiliary field $F^0$. This elimination will still maintain the action in a form that fits in the general structure (\ref{action}), where $X$ is the field $X^1$.

In order to eliminate the auxiliary field $F^0$, we first collect the terms in the action with $F^I$.
We write ${\cal L}_{1,F}$ as
\begin{equation}
{\cal L}_{1,F} =\eta_{IJ}\left(F^I+\eta^{IK}\overline{{\cal W}}_{\bar K}\right)\left(\bar F^J+\eta^{JL}{\cal W}_L\right)-
   {\cal W}_I \eta^{IJ}\overline{{\cal W}}_{\bar J}\,.
 \label{L1F}
\end{equation}
We eliminate $F^0$ and thus remain with
\begin{equation}
 {\cal L}_{1,F} \approx \left(F^1 + \overline{{\cal W}}_1\right)\left(\bar F^1 +{\cal W}_1\right)-{\cal W}_1 \overline{{\cal W}}_{\bar 1}+{\cal W}_0 \overline{{\cal W}}_{\bar 0}
 \,.
 \label{F0eliminate}
\end{equation}

The full Lagrangian ${\cal L}$, obtained after this elimination from ${\cal L}_1$ in (\ref{12XbarXD}), is then given by
\begin{equation}
e^{-1}{\cal L} = (F^1+f)(\bar F^1 + \bar f) - \bar f\, f  + \bar X^1\, A_c\, X^1 + X^1\bar B_c + B_c\bar X^1 +C_c\,,
\label{actionSC}
\end{equation}
where, according to (\ref{F0eliminate}),
\begin{equation}
f=\overline{{\cal W}}_1 = \frac{1}{3}\bar b (\bar X^{0})^2\,.\label{deff}
\end{equation}
The quantities in (\ref{actionSC}) got extra terms beyond those in (\ref{12tr0C0}) originating from the last term in (\ref{F0eliminate}). We thus obtain
\begin{eqnarray}
A_c&=&  \left[\partial _\mu \sqrt{g}g^{\mu\nu}\partial _\nu + \rmi\,e\, t_c^\mu \partial _\mu + \ft12\rmi\,\partial _\mu (e\,t_c^\mu) + e\,r_c\right] \ ,\\
r_c &=&r_0^c +{\cal W}_{01} \overline{{\cal W}}_{01}\nonumber\\
&=& -\ft16 R(\omega (e )) +\ft16\bar \psi _\mu \gamma ^{\mu \nu \rho } D^{(0)}_\nu \psi _\rho
  -A^aA_a -\ft16{\cal L}_{\rm SG,torsion}+\ft49|b X^0|^2 \,,\nonumber\\
B_c&=& B^1_c+ \overline{W}_{01}\left[{\cal W}_0\right]_{X^1=0}\nonumber\\
&& -\ft12\overline{{\cal  W}}_{IJ1}\bar \chi^{I}P_R\chi ^{J}+\frac{1}{\sqrt{2}}\overline{{\cal  W}}_{I1}\bar \psi_\mu \gamma
  ^\mu P_R \chi^I+\ft12 \overline{{\cal W} }_1\bar \psi _{\mu }P_L \gamma ^{\mu \nu }\psi _{\nu }\nonumber\\
  &=& \frac{1}{\sqrt{2}}\left[-e^{-1}\partial _\mu \left(e\bar \psi _\nu \gamma ^\mu \gamma ^\nu P_L  \chi^1\right)-\ft23\bar \chi^1 P_L \gamma ^{\mu \nu }D _\mu \psi _\nu +\rmi A^\mu \bar \psi _\mu P_L\chi^1\right]\nonumber\\
&& +\frac{1}{3}\bar b
\Bigl(2 \frac{1}{\sqrt{3}}a  (X^0)^2\bar X^0 -{\bar\chi}^0P_R\chi^0 + \sqrt{2}{\bar\psi}\cdot\gamma P_R \chi^0 \bar X^0 +
\frac{1}{2}(\bar X^0)^2 {\bar\psi}_\mu\gamma^{\mu\nu}P_L\psi_\nu\Bigr)\,,\nonumber\\
 C_c & = & -e^{-1}\bar X^0  \left[\partial _\mu \sqrt{g}g^{\mu\nu}\partial _\nu + \rmi\,e\, t_c^\mu \partial _\mu + \ft12\rmi\,\partial _\mu (e\,t_c^\mu) + e\,r_0^c\right] X^0 \nonumber\\
   &   & -\bar X^0B^0_c -X^0\bar B^0_c + C_0^c + \left[{\cal W}_0\right]_{X^1=0}\left[\overline{{\cal W}}_0\right]_{\bar X^1=0}\nonumber\\
   && +\left[ \left(-\ft12{\cal  W}_{IJ}\bar \chi^{I}P_L\chi ^{J}+\frac{1}{\sqrt{2}}\bar \psi_\mu \gamma
  ^\mu {\cal  W}_I P_L \chi^I+\ft12 \bar \psi _{\mu }P_R \gamma ^{\mu \nu }\psi _{\nu }\,{\cal W}\right)_{X^1=0} +\hc\right]\,.
 \label{Cversie1}
\end{eqnarray}

\subsection{Elimination of \texorpdfstring{$A_\mu $}{A}.}
\label{ss:elimAmu}

Then we turn to the elimination of $A_\mu $. We write as in
 \cite[(17.21)]{Freedman:2012zz}
\begin{eqnarray}
e^{-1}\frac{\delta S_1}{\delta A^\mu}
&=&\rmi\left[({\cal D} _\mu X^I)\eta_{IJ}\bar X^{\bar J}-\hc\right]+\frac{1}{2}\rmi\eta_{IJ}\bar \chi ^IP_L\gamma _\mu \chi^{\bar J}\nonumber\\
&=&2 A_\mu X^I\eta_{IJ}\bar X^J + \rmi\left[\left(\partial _\mu X^I+\frac{1}{\sqrt{2}}\bar \psi_\mu
P_L\chi^I\right)\eta_{IJ}\bar X^{\bar J}-\hc\right]\nonumber\\
&&+\frac{1}{2}\rmi\eta_{IJ}\bar \chi ^IP_L\gamma _\mu \chi^{\bar J}\,.
\end{eqnarray}
With $N$ given in (\ref{NWchoice}), we use, due to the nilpotency of $X^1$,
\begin{equation}
  \frac{1}{N}= -\frac{1}{X^0\bar X^0}-\frac{ X^1\bar X^1}{(X^0\bar X^0)^2}\,.
 \label{1overN}
\end{equation}
The solution for $A_\mu $ is
\begin{eqnarray}
  A_\mu &=& {\cal A}_\mu +{\cal A}_\mu ^{\rm F}\,,\nonumber\\
  && {\cal A}_\mu= \rmi\frac{1}{2N}\eta _{IJ}(X^I\partial _\mu \bar X^J -\bar X^I\partial _\mu X^J)={\cal A}_\mu ^0+{\cal A}_\mu ^1 \,,\nonumber\\
  &&\phantom{.}\qquad {\cal A}_\mu ^0= \rmi\frac{1}{2N}(-X^0\partial _\mu \bar X^0 +\bar X^0\partial _\mu X^0)\,,\nonumber\\
  &&\phantom{.}\qquad {\cal A}_\mu ^1= \rmi\frac{1}{2X^0\bar X^0}(-X^1\partial _\mu \bar X^1 +\bar X^1\partial _\mu X^1)\,,\nonumber\\
  && {\cal A}_\mu ^{\rm F}=\frac{1}{4N}\rmi\eta _{IJ}\left[\sqrt{2}\bar \psi _\mu(P_L\chi ^J \bar X^I-P_R\chi ^J X^I) +\bar \chi ^IP_L\gamma_\mu \chi ^J\right]={\cal A}_\mu ^{{\rm F}0}+{\cal A}_\mu ^{{\rm F}1}\,,\nonumber\\
  &&\phantom{.}\qquad {\cal A}_\mu ^{{\rm F}0}= -\frac{1}{4N}\rmi\left[\sqrt{2}\bar \psi _\mu(P_L\chi ^0\bar X^0 -P_R\chi ^0 X^0) +\ft12\bar \chi ^0\gamma _*\gamma_\mu \chi ^0\right]\,,\nonumber\\
  &&\phantom{.}\qquad {\cal A}_\mu ^{{\rm F}1}= -\frac{1}{4X^0\bar X^0}\rmi\left[\sqrt{2}\bar \psi _\mu(P_L\chi ^1\bar X^1 -P_R\chi ^1 X^1) +\ft12\bar \chi ^1\gamma _*\gamma_\mu \chi ^1\right]\,.
 \label{Amusoln}
\end{eqnarray}
After the field equations, the action contains
\begin{equation}
{\cal L}_A=e\,  N \left[{\cal A}_\mu ^0+{\cal A}_\mu ^1+{\cal A}_\mu ^{{\rm F}0}+{\cal A}_\mu ^{{\rm F}1}\right]^2\,.
 \label{actionfromAmu}
\end{equation}
Simplifications appear in the part
\begin{equation}
N \left[{\cal A}_\mu ^1+{\cal A}_\mu ^{{\rm F}1}\right]^2=\frac{1}{16N} \bar \chi^1 P_L \gamma _\mu \chi^1 \bar \chi^1 P_R \gamma ^\mu \chi^1= \frac{1}{8X^0\bar X^0} \chi ^2\bar \chi ^2\,,\qquad \chi ^2\equiv \bar \chi ^1P_L\chi^1\,,
 \label{simpleAA}
\end{equation}
where the nonzero result comes only from the square of the very last term in (\ref{Amusoln}) (after a Fierz transformation).

\subsection{Gauge fixing of the superconformal symmetry}

We now impose the gauge-fixing for dilatations and $S$-transformations
\begin{equation}
  X^0=\kappa^{-1}\sqrt{3}\,,\qquad \chi^0=0\,.
 \label{gaugefixing}
\end{equation}
Here we differ from \cite{Hasegawa:2015bza}, who took more complicated, but convenient choices in order to have the Einstein-Hilbert term $e\,R$ appearing only multiplied with $\kappa ^{-2}$. But since in this case the extra terms with $R$ are multiplied with $\chi ^2\bar \chi ^2$, we have chosen the simpler gauge fixing.

After this gauge fixing, one is left with the three fields: $(e_\mu{}^a, \psi_\mu, \chi^1)$ and the auxiliary $F=F^1$.
The expressions in (\ref{Amusoln}) simplify a lot. In fact $A_\mu ^0=A_\mu ^{{\rm F}0}=0$, and the full value of $A_\mu $ is
\begin{equation}
  A_\mu=\rmi\frac{\kappa ^2}{6}\left\{(-X^1\partial _\mu \bar X^1 +\bar X^1\partial _\mu X^1)
 -\frac{1 }{2}\left[\sqrt{2}\bar \psi _\mu(P_L\chi\bar X^1 -P_R\chi X^1) +\ft12\bar \chi \gamma _*\gamma_\mu \chi\right]\right\}\,.
 \label{Amuaftergf}
\end{equation}
Here and below we use $\chi \equiv \chi ^1$, while $\chi ^2$ is then its square as in (\ref{simpleAA}). According to (\ref{XinA}) and   (\ref{defcalA}),
\begin{equation}
  X^1 = - \frac{\chi^2}{ 2 f} (1-{\cal A})\,,\qquad
     {\cal A}= \frac{\bar \chi^2}{ 2f \bar f^2} \left(A\, \frac{\chi^2}{2f} - B\right)
  \,.
 \label{finalvalueXcalA}
\end{equation}
For the action, all what remains from (\ref{actionfromAmu}) is the part in (\ref{simpleAA}), which becomes
\begin{equation}
 e^{-1} {\cal L}_A=N \left[{\cal A}_\mu ^1+{\cal A}_\mu ^{{\rm F}1}\right]^2= \frac{\kappa ^2}{24}\chi ^2\bar \chi ^2\,.
 \label{LA}
\end{equation}
To present the result in a simpler form, we make a redefinition
\begin{equation}
  a = \kappa \, m\,, 
 \label{redefmM}
\end{equation}
and we will rewrite the dependence on $b$ in terms of (see (\ref{deff}))
\begin{equation}
 f= \kappa ^{-2}\bar b\,.
 \label{finb}
\end{equation}
We take $a$ and $b$, and thus also $m$ and $f$, real. We show in Appendix~\ref{ss:phases} that these choices amount to field redefinitions.

The expressions mentioned before simplify due to (\ref{gaugefixing}).
The full Lagrangian is at this point given by
\begin{equation}
e^{-1}{\cal L}(X,F) = (F+f)(\bar F + \bar f)   + \bar X^1\, A\, X^1 + X^1\bar B + B\bar X^1 +C\,.
\label{actionPoinc}
\end{equation}
With respect to the previous expressions, $C$ gets an extra contribution from (\ref{LA}), replacing $A_\mu $ terms, and absorbs the $(-f^2)$ term from (\ref{actionSC}). We furthermore use the explicit expressions in (\ref{Wderiv}) and the redefinitions  (\ref{redefmM}), (\ref{finb}).
The quantities in this expression are
\begin{eqnarray}
 A &=&   \Box + \rmi t^\mu \partial _\mu + \ft12\rmi e^{-1}\partial _\mu (e t^\mu) + r\,, \qquad
  \Box =\frac{1}{\sqrt{g}}\partial _\mu \sqrt{g}g^{\mu\nu}\partial _\nu\,,\nonumber\\
   t^\mu &=&\ft14 {\rmi} {\bar \psi}_\nu \gamma_\star \gamma^{\nu\rho\mu}\psi_\rho= -\ft14e^{-1}\varepsilon ^{\mu \nu \rho \sigma }\bar \psi _\nu \gamma _\rho \psi _\sigma \,,\nonumber\\
 r & = & -\ft16 \left[R(\omega (e )) -\bar \psi _\mu \gamma ^{\mu \nu \rho } D^{(0)}_\nu \psi _\rho
   +{\cal L}_{\rm SG,torsion}
   -8\kappa ^2\,f ^2\right]  \,, \nonumber\\
B  &=&   \frac{1}{\sqrt{2}}\left[-e^{-1}\partial _\mu \left(e\bar \psi _\nu \gamma ^\mu \gamma ^\nu P_L  \chi\right)-\ft23\bar \chi P_L \gamma ^{\mu \nu }D _\mu \psi _\nu \right]
 + f \Bigl(2 m + \frac{1}{2} {\bar\psi}_\mu\gamma^{\mu\nu}P_L\psi_\nu\Bigr)\,,\nonumber\\
C&=& 
\frac{1}{2\kappa ^2} \left[ R(\omega (e )) -\bar \psi _\mu \gamma ^{\mu \nu \rho } D^{(0)}_\nu \psi _\rho
  +{\cal L}_{\rm SG,torsion} \right] +3\frac{m^2}{\kappa ^2} -f ^2 \nonumber\\
&&+\frac{1}{\sqrt{2}}f \bar \psi_\mu \gamma
  ^\mu   \chi+\frac{m}{2\kappa ^2}\bar \psi _{\mu } \gamma ^{\mu \nu }\psi _{\nu } +\frac{\kappa ^2}{24}\chi ^2\bar \chi ^2\nonumber\\
 &&-\ft12\bar \chi  \slashed{D}^{(0)}\chi -\ft{1}{32}\rmi\,e^{-1}\varepsilon ^{\mu \nu \rho \sigma }\bar \psi _\mu \gamma _\nu \psi _\rho \bar \chi \gamma _*\gamma _\sigma \chi-\ft12\bar \psi _\mu P_R\chi \bar \psi ^\mu P_L\chi\,,
 \label{finalrBC}
\end{eqnarray}
where $D_\mu \psi _\nu $ is given in (\ref{covder0}).

\subsection{Elimination of \texorpdfstring{$F^1$}{F}.}

We can now apply Sec. \ref{ss:solnF} and obtain
\begin{eqnarray}
 e^{-1}{\cal L} & = &\frac{1}{2\kappa ^2} \left[ R(\omega (e )) -\bar \psi _\mu \gamma ^{\mu \nu \rho } D^{(0)}_\nu \psi _\rho
  +{\cal L}_{\rm SG,torsion} \right] +3\frac{m^2}{\kappa ^2} -f ^2 \nonumber\\
&&+\frac{1}{\sqrt{2}}f \bar \psi_\mu \gamma
  ^\mu   \chi+\frac{m}{2\kappa ^2}\bar \psi _{\mu } \gamma ^{\mu \nu }\psi _{\nu } +\frac{\kappa ^2}{24}\chi ^2\bar \chi ^2\nonumber\\
 &&-\ft12\bar \chi  \slashed{D}^{(0)}\chi -\ft{1}{32}\rmi\,e^{-1}\varepsilon ^{\mu \nu \rho \sigma }\bar \psi _\mu \gamma _\nu \psi _\rho \bar \chi \gamma _*\gamma _\sigma \chi-\ft12\bar \psi _\mu P_R\chi \bar \psi ^\mu P_L\chi\nonumber\\
&& +\frac{\bar  \chi^2}{2 f }A \frac{\chi^2}{2 f } -\left(\frac{ \chi^2}{2 f } \bar B+\frac{\bar \chi^2}{2 f } B\right) - \frac{\chi^2\bar \chi^2}{16 f ^4}\left(\frac{\Box\chi^2}{f }-2B\right)\left(\frac{\Box\bar \chi^2}{f }-2\bar B\right)\,,\label{finalactionfA}
\end{eqnarray}
where $D^{(0)}$ and ${\cal L}_{\rm SG,torsion}$ are given in (\ref{covder0}), and the quantities in (\ref{finalrBC}) are used. This is the result that was given in \cite{Bergshoeff:2015tra}.

\subsection{Transformation laws}

%

When using the gauge condition (\ref{gaugefixing}), the decomposition law for the $S$-supersymmetry is as in \cite[(16.46)]{Freedman:2012zz}:
\begin{equation}
  P_L\eta = \frac12\rmi P_L\slashed{A}\epsilon -\frac{\kappa
  }{2\sqrt{3}}F^0P_L\epsilon \,.
\label{decomppureN1}
\end{equation}
Therefore, the transformation laws are
\begin{eqnarray}
\delta e^a_\mu &=& \frac12 \bar\epsilon \gamma^a\psi_\mu \ ,\nonumber\\
\delta P_L\psi_\mu &=& P_L\bigg(\partial_\mu +\frac14 \omega_{\mu ab}(e,\psi)\gamma^{ab}-\frac32iA_\mu +\frac12i\gamma_\mu \slashed{A} +
\frac{\kappa}{2\sqrt3}\gamma_\mu \bar F^0 \bigg)\epsilon\label{deltpsi}\,,
\end{eqnarray}
 with
\begin{equation}
  F\ \rightarrow \ F^0=\overline{{\cal W}}_0= \sqrt{3}\,\frac{m}{\kappa } + \frac{2}{\sqrt{3}}\kappa f \, X^1= \sqrt{3}\,\frac{m}{\kappa }- \frac{1}{\sqrt{3}}\kappa \chi ^2(1-{\cal A})\,,
 \label{Fvalue}
\end{equation}
and $A_\mu$  as in (\ref{Amuaftergf}).
The transformation of the fermion follows from \cite[(16.33)]{Freedman:2012zz} and
\begin{eqnarray}
 \delta P_L\chi  &=& \frac1{\sqrt{2}} P_L\left(\slashed{\cal D} X^1 +
F^1\right)\epsilon +\sqrt{2} X^1 P_L\eta \,, \nonumber\\
&=&- \frac{f}{\sqrt{2}} P_L\epsilon+\frac1{\sqrt{2}} P_L\left[\slashed{\cal D} X^1
-  f {\cal A} \left(1-3\bar{{\cal A}}-\frac{\chi^2}{2f ^3}\bar B\right)
\right]\epsilon +\sqrt{2} X^1 P_L\eta \,,
\label{transfoOmega}
\end{eqnarray}
where
\begin{eqnarray}
 {\cal D}_\mu  X^1 &=&\partial _\mu  X^1- \rmi \, A_\mu
  X^1 -\frac{1}{\sqrt{2}}\bar \psi _\mu P_L\chi \,, \nonumber\\
F^1 &=& -  f \left[1+{\cal A} \left(1-3\bar{{\cal A}}-\frac{\chi^2}{2f ^3}\bar B\right)\right]\,.
\end{eqnarray}
This further simplifies since
\begin{equation}
  A_\mu X^1 =0\,,\qquad F^0 X^1=\sqrt{3}\,\frac{m}{\kappa } X^1\,.
 \label{simpleAXFX}
\end{equation}
The transformation law is then
\begin{equation}
 \delta P_L\chi =\frac1{\sqrt{2}} P_L\left[-f+(\slashed{\partial }-m)X^1
-  f {\cal A} \left(1-3\bar{{\cal A}}-\frac{\chi^2}{2f ^3}\bar B\right) \right]\epsilon
-\frac{1}{2}P_L\gamma ^\mu \epsilon \bar \psi _\mu P_L\chi
\,.
\label{transfoOmegasimple}
\end{equation}

\section{Comments}
\label{ss:comments}

We have given here the details of the result explained in \cite{Bergshoeff:2015tra}. The final Lagrangian (\ref{finalactionfA}) has local supersymmetry given by (\ref{deltpsi}) and (\ref{transfoOmegasimple}). There are two independent constants, $f$ and $m$, which give respectively positive and negative contribution to the energy.  Therefore we can have supergravities with dS, Minkowski and AdS vacua. The parameter $f$ is the supersymmetry breaking parameter. Thus, in agreement with the general algebraic arguments repeated in Sec. \ref{ss:algebrasdS}, de Sitter vacua are only possible in case of supersymmetry breaking.

In \cite{Bergshoeff:2015tra} also other features of this action were discussed. It was also clarified how one can define a unitary supersymmetry gauge $\chi =0$, in which case the action becomes very simple.

Meanwhile, the result has been extended to couplings with matter multiplets \cite{Hasegawa:2015bza,Kallosh:2015sea,Kallosh:2015tea,Schillo:2015ssx,Bandos:2015xnf,Kallosh:2015pho}, and other similar constraints of multiplets in supergravity have been considered \cite{Antoniadis:2014oya,Dudas:2015eha,Ferrara:2015gta,Kuzenko:2015yxa,Antoniadis:2015ala,Ferrara:2015cwa,Hasegawa:2015era,Dall'Agata:2015zla,Kahn:2015mla,Ferrara:2015tyn,Kuzenko:2015rfx,Dall'Agata:2015lek}.

\medskip
\section*{Acknowledgments}

\noindent We are grateful to Timm Wrase for interesting discussions on this work. We acknowledge hospitality in Texas A\&M  and in the university of  Utrecht.
The work of A.V.P. is supported in part by the FWO - Vlaanderen, Project No.
G.0651.11, and in part by the Interuniversity Attraction Poles Programme
initiated by the Belgian Science Policy (P7/37) and in part by COST Action MP1210
''The String Theory Universe''.

\appendix

\section{Equivalence of using field equations before or after constraints}
\label{app:feLagrM}

We consider an action with Lagrange multipliers $\lambda^i$, fields $X^\alpha$ that will be solved by the constraints, and the other fields $F^a$:
\begin{equation}
  S(F^a,X^\alpha,\lambda^i)=S_1(F^a,X^\alpha) + \lambda^i C_i(F^a,X^\alpha)\,.
 \label{LwithLagrmult}
\end{equation}
We are using the condensed DeWitt notation so that the sum over $i$ in the last term includes an integral over spacetime.
We suppose that there exist an expression
\begin{equation}
  X^\alpha =x^\alpha(F^a)\,,
 \label{solveconstraints}
\end{equation}
that solves the constraints, i.e.
\begin{equation}
  C_i(F^a,x^\alpha(F^a))=0\,.
 \label{Csolved}
\end{equation}

We will now show that such a solution of the constraints, solving also the other field equations of (\ref{LwithLagrmult}), should be a solution of the effective Lagrangian
\begin{equation}
  S_{\rm eff} (F^a)= S_1(F^a,x^\alpha(F^a))\,.
 \label{Lsolved}
\end{equation}

The field equations from (\ref{LwithLagrmult}) are
\begin{eqnarray}
\frac{\delta S}{\delta \lambda^i}&=&C_i=0\,,\nonumber\\
 \frac{\delta S}{\delta F^a} & = & \frac{\delta S_1}{\delta F^a}+ \lambda^i\frac{\delta C_i}{\delta F^a}=0\,, \nonumber\\
 \frac{\delta S}{\delta X^\alpha} & = & \frac{\delta S_1}{\delta X^\alpha}+ \lambda^i\frac{\delta C_i}{\delta X^\alpha}=0\,.
 \label{feL}
\end{eqnarray}
The first ones are solved by (\ref{solveconstraints}), whose derivative w.r.t. $F^a$ implies
\begin{equation}
  \frac{\delta C_i}{\delta F^a}+\frac{\delta C_i}{\delta X^\alpha}\frac{\delta x^\alpha}{\delta F^a}=0\,.
 \label{derivCsolved}
\end{equation}
We now assume that we have a solution of the other two equations in (\ref{feL}). Then the second equation, using (\ref{derivCsolved}), implies
\begin{equation}
  \frac{\delta S_1}{\delta F^a} -\lambda^i\frac{\delta C_i}{\delta X^\alpha}\frac{\delta x^\alpha}{\delta F^a}=0\,.
 \label{conseqfeL2}
\end{equation}
Combining this with the third equation in (\ref{feL}) leads to
\begin{equation}
  \frac{\delta S_{\rm eff}}{\delta F^a}=\frac{\delta S_1}{\delta F^a}+ \frac{\delta S_1}{\delta X^\alpha}\frac{\delta x^\alpha}{\delta F^a}=0\,.
 \label{feLs}
\end{equation}
Thus it must be a solution of the field equations effective action. This solves  in general a problem raised in \cite[Appendix C]{Kuzenko:2011tj}.

In our toy model this goes as follows. In the field equation for $F$,  (\ref{feF}), we use the derivative of the constraint (\ref{feLambda}) with respect to $F$, which is $X+F\frac{\partial X}{\partial F}=0$. This leads to
\begin{equation}
0=  \bar F + f - 2 \Lambda F\frac{\partial X}{\partial F}=\bar F + f + 2 \Lambda X\,.
 \label{feFstep2}
\end{equation}
We now use $\Lambda$ as obtained from (\ref{feX}), but since $\Lambda $ is multiplied in (\ref{feFstep2}) by $X$, we only have to use one term and obtain
\begin{equation}
0=  \bar F + f -  \frac{1}{F}X\Box \bar X\,,
 \label{feFstep3}
\end{equation}
which is the field equation that follows from (\ref{effectiveL}).

\section{Details of calculations of the action}
\label{app:detailsaction}

In this appendix we write the expression of  (\ref{DbecomesF}) in a form convenient for  (\ref{action}), which can then as well be used for $X=X^0$ as for $X=X^1$.
Our starting point is  (\ref{fullactionFform}). However,  (\ref{XL1fesimpler}) is more useful to obtain $AX+B$.
We further combine results mentioned in Chapter 16 and 17 of \cite{Freedman:2012zz}, leading to
\begin{eqnarray}
 [X\bar X]_D=\ft12 [X\bar F]_F & = & \bar X  \left[\partial _\mu \sqrt{g}g^{\mu\nu}\partial _\nu + \rmi e\, t^\mu \partial _\mu + \ft12\rmi\partial _\mu(e\, t^\mu) +e\, r_0\right] X
  \nonumber\\
   &   & +e\, \bar X\, B_0 + e\,X\bar B_0 +e \, C_0\,,
 \label{XbarXD}
\end{eqnarray}
(up to total derivatives) where
\begin{eqnarray}
t^\mu &=&-2 A^\mu +\ft14 {\rmi} {\bar \psi}_\nu \gamma_\star \gamma^{\nu\rho\mu}\psi_\rho\,,\nonumber\\
r_0 &=&-\ft16 R(\omega (e,\psi )) +\ft16\bar \psi _\mu \gamma ^{\mu \nu \rho } D_\nu \psi _\rho
  -A^aA_a  \,,\nonumber\\
 B_0 &=& \frac{1}{\sqrt{2}}\left(-\bar \psi _\mu \gamma ^\nu \gamma ^\mu P_L D_\nu \chi -D_\mu \bar \psi ^\mu P_L\chi  +\ft13 \bar \chi P_L \gamma ^{\mu \nu }D _\mu \psi _\nu
 +\rmi A^\mu \bar \psi _\mu P_L\chi\right.\nonumber\\ &&\phantom{\frac{1}{\sqrt{2}}\ }\left.+\bar \psi \cdot \gamma \psi _\nu \bar \psi ^\nu P_L\chi \right)\,,
 \nonumber\\
 C_0&=&F\bar F  -\ft12\bar \chi  \slashed{D}\chi+\ft14\rmi \bar \chi \gamma _*\gamma ^\mu \chi A_\mu -\ft14\left(\bar \chi P_L\gamma ^\mu \gamma ^\nu \psi _\mu \bar \psi _\nu P_R\chi +\hc\right) \,.
 \label{tr0C0}
\end{eqnarray}
Here
\begin{equation}
  D_\mu \chi =\left(\partial _\mu +\ft14 \omega _\mu {}^{ab}(e,\psi )\gamma _{ab}\right)\chi  \,,\qquad D_\mu \psi _\nu =\left(\partial _\mu +\ft14 \omega _\mu {}^{ab}(e,\psi )\gamma _{ab}\right) \psi _\nu\,.
 \label{korteD}
\end{equation}

We give here a few remarks on how the expression (\ref{XbarXD}) was obtained.
To find the expression $t_\mu $, we extract the terms with a derivative on $\bar X$ from (\ref{XL1fesimpler}). Except for the $\omega (e)$ terms that are already included in the $\Box$ part of $A$, there are the terms with $A_\mu $ and quadratic gravitino terms. The former lead in an obvious way to a contribution to $t^\mu $ proportional to $A^\mu$, while there are two quadratic gravitino contributions. One originates from the $\omega _{\mu ab}$ in the first line of (\ref{covderchiralmult}), for which we use \cite[(17.167)]{Freedman:2012zz}, and the other from the ${\cal D}'\chi $ term in (\ref{XL1fesimpler}). These coefficients of $e\partial _\rho \bar X$ add up to
\begin{equation}
  \ft12\bar \psi\cdot \gamma\psi^\rho -\ft12\bar \psi_\mu P_R\gamma^{\mu\nu}\gamma^\rho\psi_\nu=\ft14\bar \psi_\mu \gamma_*\gamma^{\mu\nu\rho}\psi_\nu\,.
 \label{trhofermionic}
\end{equation}
This leads to the expression for $t^\mu $ in (\ref{tr0C0}).
The quantity $r_0$ in  (\ref{tr0C0}) is obtained from the calculation of \cite[(16.44)]{Freedman:2012zz}, where the $X\bar X$ in (\ref{XbarXD}), which is $Z\bar Z$ in \cite[Ch.16]{Freedman:2012zz} is replaced by $-3\kappa ^{-2}$ according to \cite[(16.40)]{Freedman:2012zz}. In the expression for $C_0$ the last terms follow from expanding the covariant derivative ${\cal D}_\mu \chi $.

Now we write out the torsion terms. Extracting these in the expression of $r_0$ from the scalar curvature and the covariant derivative of the gravitino produces the well-known supergravity torsion term: $-\ft16{\cal L}_{\rm SG,torsion}$.

For $C_0$ we find
\begin{eqnarray}
  C_0&=& F\bar F  -\ft12\bar \chi  \slashed{D}^{(0)}\chi+\ft14\rmi \bar \chi \gamma _*\gamma ^\mu \chi A_\mu + C_{0\,{\rm t}}\nonumber\\
  C_{0\,{\rm t}}&=&-\ft1{32}\bar \chi \gamma ^{\mu\nu \rho }\chi \bar \psi _\mu \gamma _\nu \psi _\rho -\ft14\left(\bar \chi P_L\gamma^{\mu \nu }\psi _\mu \bar \psi _\nu P_R\chi +\hc\right)-\ft12 \bar \chi P_L\psi _\mu \bar \psi^\mu P_R\chi \nonumber\\
&=&-\ft1{32}\bar \chi \gamma ^{\mu\nu \rho }\chi \bar \psi _\mu \gamma _\nu \psi _\rho +\ft1{16}\bar \psi _\nu \gamma ^\rho \psi _\mu \left(\bar \chi P_L\gamma ^{\mu \nu }\gamma _\rho \chi +\hc\right)
%
%
-\ft12 \bar \chi P_L\psi _\mu \bar \psi^\mu P_R\chi\nonumber\\
&=&  - \ft1{32}\bar \chi \gamma ^{\mu\nu \rho }\chi \bar \psi _\mu \gamma _\nu \psi _\rho +\ft1{16}\bar \psi _\nu \gamma _\rho \psi _\mu \bar \chi \gamma ^{\mu \nu \rho }\chi
-\ft12 \bar \chi P_L\psi _\mu \bar \psi^\mu P_R\chi\nonumber\\
&=&   -\ft1{32}\rmi\,e^{-1}\,\varepsilon ^{\mu \nu \rho \sigma }\bar \psi _\mu \gamma _\nu \psi _\rho \bar \chi \gamma _*\gamma _\sigma \chi-\ft12\bar \psi _\mu P_R\chi \bar \psi ^\mu P_L\chi  \,,
 \label{C0torsion}
\end{eqnarray}
where $D_\mu ^{(0)}$ is the derivative without torsion, and we used symmetry properties of fermion bilinears.

The torsion terms in $B_0$ (taking into account that $D_\mu \bar \psi ^\mu$ contains $\Gamma _{\mu \nu }^\mu \bar \psi ^\nu $) combine to
\begin{equation}
  B_0= \frac{1}{\sqrt{2}}\left[-e^{-1}\partial _\mu \left(e\bar \psi _\nu \gamma ^\mu \gamma ^\nu P_L  \chi\right)-\ft23\bar \chi P_L \gamma ^{\mu \nu }D _\mu \psi _\nu +\rmi A^\mu \bar \psi _\mu P_L\chi\right]\,.
 \label{B0torsion}
\end{equation}
Note that here the torsion terms are still included in $D _\mu \psi _\nu $.  In fact we have
\begin{equation}
  -\ft23 P_L \gamma ^{\mu \nu }D _\mu \psi _\nu= -\ft23 P_L \gamma ^{\mu \nu }D^{(0)} _\mu \psi _\nu +\ft16 P_L\left(\psi _\mu \bar \psi \cdot \gamma \psi _\mu
   +\ft1{2}\gamma ^{\mu \nu }\psi _\mu \bar \psi \cdot \gamma \psi _\nu -\ft1{4}\gamma ^{\mu \nu }\psi _\rho \bar \psi_\mu   \gamma^\rho  \psi _\nu\right)\,.
 \label{torsioninB0}
\end{equation}

\section{Phases of the constants}
\label{ss:phases}
We prove here that the phases of the constants $a$ and $b$ in (\ref{NWchoice}) are irrelevant, which means that they can be removed by field redefinitions.
Suppose
\begin{equation}
  a=|a|\rme^{\rmi\theta _a}\,,\qquad b=|b|\rme^{\rmi\theta _b}\,.
 \label{phasesab}
\end{equation}
In the conformal variables we can replace
\begin{equation}
  X^0=X^0_{(R)}\rme^{-\rmi\theta _a/3}\,,\qquad X^1=X^1_{(R)}\rme^{\rmi(2\theta _a/3-\theta _b)}\,,
 \label{firstphasesX}
\end{equation}
such that after these replacement, the formulas with complex $a$ and $b$ are the same as those with $|a|$ and $|b|$, using $X^0_{(R)}$ and $X^1_{(R)}$ in stead of $X^0$ and $X^1$.
We then have to do the same phase transformation for the other fields in the multiplets. E.g.
\begin{equation}
 P_L \chi ^1= P_L\chi ^1_{(R)}\rme^{\rmi(2\theta _a/3-\theta _b)}\,,\qquad F^1= F^1_{(R)}\rme^{\rmi(2\theta _a/3-\theta _b)}\,.
 \label{firstphaseschiF}
\end{equation}

This is consistent for all formulas until we go to the Poincar\'{e} gauge (\ref{gaugefixing}). Using this with the replacement (\ref{firstphasesX}) would lead to another gauge fixing for the $R$-symmetry $\U(1)$. Therefore, we have to compensate for this difference by a chiral rotation.
We perform a compensating chiral rotation with $\lambda _T=\theta _a/3$
 (see the weights of the fields in \cite[Table 17.1]{Freedman:2012zz} with $w=1$ for the chiral multiplet). Also the gravitino has to be redefined. The final redefinitions are
\begin{eqnarray}
 P_L\psi _\mu  & = & P_L\psi _{\mu (R)} \rme^{\rmi\theta _a/2}\,,\nonumber\\
 X^0&=& X^0_{(R)} \,,\nonumber\\
 X^1 & = & X^1_{(R)} \rme^{\rmi(\theta _a-\theta _b)}\,,\nonumber\\
 P_L\chi ^1 &=& P_L\chi ^1_{(R)}\rme^{\rmi(\theta _a/2-\theta _b)}\,,\nonumber\\
 F^1&=&F^1_{(R)}\rme^{-\rmi\theta _b}\,.
 \label{redeffinal}
\end{eqnarray}
E.g., using the definition in (\ref{deff}) (see also  (\ref{redefmM}) and (\ref{finb}))
\begin{equation}
 m=|m|\rme^{\rmi\theta _a}\,,\qquad   f=|f|\rme^{-\rmi\theta _b}\,.
 \label{fmodulus}
\end{equation}
These replacements do not modify the quantity $A$: i.e. $A=A_{(R)}$, where  the latter means that we replace all fields with their $(R)$ values. But $B$ undergoes a chiral rotation: $B=B_{(R)}\rme^{\rmi(\theta _a-\theta _b)}$.
On the other hand, the quantity ${\cal A}$ is invariant, which is consistent with  (\ref{finalvalueXcalA}).

This shows that there is no physics in the phases of the constants $a$ and $b$ (or $f$ and $m$).


\providecommand{\href}[2]{#2}\begingroup\raggedright\endgroup

\end{document}